\documentclass[twocolumn]{autart} 
\usepackage{natbib}
\usepackage{har2nat}
\usepackage{amsmath}
\usepackage{amssymb}
\usepackage{bbm}
\usepackage{color}
\usepackage{epstopdf}
\usepackage{siunitx}
\usepackage{bm}
\usepackage{graphicx}
\usepackage{subfig}
\usepackage{enumerate}
\usepackage{wrapfig}
\usepackage{algorithmicx,algorithm}
\usepackage[noend]{algpseudocode}

\def\and{\text{and}}

\newcommand{\mathletter}[1]{%
	\expandafter\newcommand\csname b#1\endcsname{\mathbb #1}
	\expandafter\newcommand\csname c#1\endcsname{\mathcal #1}
	\expandafter\newcommand\csname f#1\endcsname{\mathfrak #1}
	\expandafter\newcommand\csname til#1\endcsname{\widetilde #1}
	\expandafter\newcommand\csname ha#1\endcsname{\widehat #1}
	\expandafter\newcommand\csname s#1\endcsname{\boldsymbol #1}
}%
\def\mathletters#1{\mathlettersB #1,,}
\def\mathlettersB#1,{\ifx,#1,\else\mathletter #1\expandafter\mathlettersB\fi}
\mathletters{A,B,C,D,E,F,G,H,I,J,K,L,M,N,O,P,Q,R,S,T,U,V,W,X,Y,Z}

\def \qed {\hfill \vrule height6pt width 6pt depth 0pt}
\def\bee{\begin{equation}}
	\def\ene{\end{equation}}
\def\been{\begin{equation*}}
	\def\enen{\end{equation*}}
\def\beq{\begin{eqnarray}}
	\def\enq{\end{eqnarray}}
\def\bmatri{\begin{bmatrix}}
	\def\ematri{\end{bmatrix}}

\newtheorem{defi}{Definition}
\newtheorem{theo}{Theorem}
\newtheorem{lemma}{Lemma}
\newtheorem{coro}{Corollary}

\newtheorem{example}{Example}

\def\mP{{\mathcal P}}
\def\mD{{\mathcal D}}
\def\mL{{\mathcal L}}

\begin{document}
	\begin{frontmatter} 
		\title{Minimum Input Design for Direct Data-driven Property Identification of Unknown Linear Systems} 
		\thanks[footnoteinfo] {This work is supported by the National Natural Science Foundation of China under grant No. 62033006. The material in this paper was not presented at any conference.} 
		\author{Shubo~Kang}, 
		\ead{ksb20@mails.tsinghua.edu.cn}
		\author{Keyou~You\corauthref{cor}}
		\corauth[cor]{Corresponding author} 
		\ead{youky@tsinghua.edu.cn}
		\address{Department of Automation, and Beijing National Research Center for Information Science and Technology, Tsinghua University, Beijing 100084, China.}
	
	\begin{abstract} 
		In a direct data-driven approach, this paper studies the {\em property identification(ID)} problem to analyze whether an unknown linear system has a property of interest, e.g., stabilizability and structural properties. In sharp contrast to the model-based analysis, we approach it by directly using the input and state feedback data of the unknown system. Via a new concept of sufficient richness of input sectional data, we first establish the necessary and sufficient condition for the minimum input design to excite the system for property ID. Specifically, the input sectional data is sufficiently rich for property ID {\em if and only if} it spans a linear subspace that contains a  property dependent minimum linear subspace,  any basis of which can also be easily used to form the minimum excitation input. Interestingly, we show that  many structural properties  can be identified with the minimum input  that is however unable to identify the explicit system model. Overall, our results rigorously quantify the advantages of the direct data-driven analysis over the model-based analysis for linear systems in terms of data efficiency.
		
	\end{abstract}

	\end{frontmatter}

	\section{Introduction}
		The modern control theory has been firmly rooted in tradition of model-based methods, which inarguably requires the system modeling and parameter identification (ID). In this paper, we are motivated by two questions that possibly promote rethinking of this foundation: (a) whether the parameter ID is indispensable to our control theory, and (b) if not, can we address control problems in a data-driven fashion as in machine learning  that directly takes data samples to make decisions. To this end, this work focuses on the {\em property ID} problem to analyze whether an unknown linear system has a specified property of interest in a direct data-driven approach with minimum input data. 
	
	In sharp contrast to model-based methods, we directly use the input and state feedback data to express our property ID results. This agrees with the spirit to learn an optimal policy from data in reinforcement learning \citep{sutton2018reinforcement}.  In fact, research on directly learning system property or control policy from data has a long history. See an early survey \citep{fu1970learning} and references therein. Recently,  the direct data-driven methods have regained significant attention and the comparisons with model-based methods are well documented in quite a few excellent works \citep{hou2013, tu2019gap, recht2019tour, baggio2021data,dorfler2022bridging}.
	
		Specifically, \citet{coulson2019data} study direct data-driven control  using behavioral system theory  where the system dynamics is expressed by a series of  input and output data. Under the persistently exciting (PE) condition,  some data-enabled formulas are established to solve various control design problems \citep{coulson2019data,de2019formulas, berberich2020data,guo2021data, markovsky2021behavioral, coulson2021distributionally}. Though such a control technique, e.g., DeePC  \citep{berberich2020data,coulson2021distributionally}, shows very good performance in many difficult tasks, it is known from  \citet{ljung1999system} that the PE condition is also sufficient for the parameter ID. From this perspective, these works implicitly assume that the full system model can be identified and thus do not touch the minimum input design issue. 
		
	In \citet{fazel2018global}, the policy optimization (PO) method has been proposed to directly use data to learn an optimal control policy for the classical Linear Quadratic Regulator (LQR) problem.  After establishing its optimization landscape, they rigorously provide the sample complexity of the direct data-driven PO method. In fact, classical control problems with unknown dynamics have re-attracted the interest from the data-driven point of view, including the LQG control \citep{tang2021analysis}, $\mathcal{H}_2/\mathcal{H}_{\infty}$ control \citep{zhang2019policy}, risk-constrained LQR control \citep{zhao2021global}, and their variants \citep{mohammadi2021convergence}.  These works not only strengthen our understanding of classical control problems, but also serve as ideal benchmarks to evaluate the data efficiency of reinforcement learning algorithms.  However, they explicitly assume that the unknown system is stabilizable or a stabilizing gain is given as prior. We show in this work that such an assumption is non-trivial as verifying stabilizability of an unknown linear system is almost as difficult as identifying system parameters.

	Fundamentally different from the aforementioned works, we study the  property ID problem in a direct data-driven approach and aim to quantify its  advantages over the model-based methods. 
	 To achieve it, one of our central idea lies in the proposal of a novel concept of {\em sufficient richness of the input sectional data} (cf. Definition \ref{def_sr}). As in the system ID \citep{ljung1999system}, sufficient richness formalizes our minimum requirement on the sectional input data to excite the system for property ID.  That is, we cannot complete the property ID if the input data is not sufficiently rich.  What looks unusual is the so-called input sectional data which consists of {\em multiple} pairs of initial state and control input, since  the initial state is also an external input to excite the system and can be potentially designed. Though it may not be trivial to design the initial state for any unknown system,  we can indeed achieve it for many systems, e.g., inverted pendulum systems.
	
	Our major result shows that an input sectional data is sufficiently rich for the property ID of an unknown linear system {\em if and only if} it spans a linear subspace that contains a property-dependent minimum linear subspace.  For a trivial case, the minimum linear subspace of the parameter identifiability is $\bR^{m+n}$ where $n$ and $m$ are the dimensions of the state and control input vectors, respectively. Such a property dependent minimum linear subspace not only decide the minimum number of input sectional data for property ID, which equals the number of its basis vectors, but also show how to easily design the input via its basis.  Moreover, it enables to rigorously quantify the advantage of data-driven approach for system analysis over the model-based one in terms of data efficiency.  Specifically, if this minimum subspace is  {\em strictly} contained in $\bR^{m+n}$, one can only achieve the direct data-driven property ID with its minimum input that is however impossible to identify system parameters.  
		
	Then, a follow-up question is how to explicitly depict the above minimum subspace. Firstly, we study the two most important properties -- stabilizability and controllability -- of unknown linear systems. Except for the controllability of scalar systems, the minimum linear subspace is of the same as that of identifiability, meaning that  the input sectional data for property ID is also sufficiently rich for system ID. Note that it is still possible to identify these properties before being able to identify system parameters if the feedback data is ``good" enough. From this point of view, the data-driven property ID is still more data efficient than the model-based approach.   
	
	Secondly, we proceed to identify other structural properties of system matrices, including their sparsity patterns and some complex combinations of linearly constrained structures.  A noticeable example is on identifying the interconnections between state variables.  In these  cases, we explicitly characterize their minimum linear subspaces and reveal that there are quite a few structural properties that can be identified in a direct data-driven approach using the minimum input  that is unable to identify system parameters. Finally, we establish data-enabled necessary and sufficient conditions for direct data-driven property ID with sufficiently rich input. 
	
We note that the direct data-driven property ID can be traced back to the pioneer work \citep{park2009stability}, where a condition on data space for stabilizability of linear systems is given.  Subsequently, data conditions on both input and output data are established  for controllability and observability, \citet{wang2011data,liu2014data,zhou2018data,mishra2020data,van2020data},  dissipativity \citep{maupong2017lyapunov,koch2021determining}, as well as their extensions to nonlinear systems \citep{martin2021data}.  However, all these works do not consider the input design, whose importance has well acknowledged in the parameter ID via the concept of PE \citep{ljung1999system}.  In this work, we use the property-dependent minimum linear subspace to achieve this goal. 
	
	The rest of this paper is organized as follows. In Section \ref{sec_pf}, we describe our problem of interest. In Section \ref{sec_mini}, we derive the necessary and sufficient condition of the sufficient richness. In Section \ref{sec_sr}, we give specific conditions for identifying different fundamental properties, e.g. stabilizability, controllability and structural properties. In Section \ref{sec_dd}, we show how to achieve the direct data-driven property ID. Finally, Section \ref{sec_dc} contains the conclusions and discussions on the future works. 
	
	\section{Problem formulation} \label{sec_pf}
	Consider a discrete linear time-invariant system 
	\begin{equation}
		\label{linearsys}
		\boldsymbol{x}_{t+1} = A^*\boldsymbol{x}_t+B^*\boldsymbol{u}_t,
	\end{equation}
	where $\boldsymbol{x} \in \bR^n$ is the state vector and $\boldsymbol{u} \in \bR^m$ is the control input vector. If the explicit model $(A^*,B^*)$ is known, the model-based linear system theory \cite{trentelman2012control} shows us how to verify whether the linear system \eqref{linearsys} has some specified properties of interest, e.g., controllability and stabilizability, which we refer to as the {\em property identification (ID)} problem in this work. An interesting question is whether the explicit model is necessary for the property ID of linear systems. If not, can we achieve it in a direct data-driven fashion with minimum input design? Such a problem is complicated as illustrated by the following examples. 
	
	\begin{example}(Stabilizability) \label{firstexample}
		We consider an unknown linear system of \eqref{linearsys} with $n=2$ and $m=1$, and aim to identify its stabilizability in a direct data-driven fashion. Suppose that we have excited the system with the following two pairs of control-(initial) state inputs, i.e.,
		\bee
		\label{input}
		u_0^{(1)}=-1,\ \boldsymbol{x}_0^{(1)}=\begin{bmatrix}1\\0\end{bmatrix};
		u_0^{(2)}=-1,\ \boldsymbol{x}_0^{(2)}=\begin{bmatrix}0.5\\1\end{bmatrix}	
		\ene
		where the superscript $^{(\cdot)}$ denotes the excitation index.  
		Regardless of the values of the state  feedback $\boldsymbol{x}_1^{(1)}$ and $\boldsymbol{x}_1^{(2)}$, it is clearly impossible to identify the explicit model $(A^*,B^*)$ using the above input and state feedback data. 
		
		However, if $\boldsymbol{x}_1^{(1)}=[0.5,1]'$ and $\boldsymbol{x}_1^{(2)}=[-1/4,1]'$,  it follows from \citet{van2020data} that the unknown linear system in \eqref{linearsys} can still be stabilized via a static state feedback with the following gain matrix 
		\bee
		\label{fbgain}
		K=[u_0^{(1)},u_0^{(2)}][\boldsymbol{x}_0^{(1)},\boldsymbol{x}_0^{(2)}]^{-1}.
		\ene
		Specifically, one can easily verify that the closed-loop system matrix
		\bee
		\begin{split}
		&A^*+B^*K\\
		&=(A^*[\boldsymbol{x}_0^{(1)},\boldsymbol{x}_0^{(2)}]+B^*[u_0^{(1)},u_0^{(2)}])[\boldsymbol{x}_0^{(1)},\boldsymbol{x}_0^{(2)}]^{-1}\\
		&=[\boldsymbol{x}_1^{(1)},\boldsymbol{x}_1^{(2)}][\boldsymbol{x}_0^{(1)},\boldsymbol{x}_0^{(2)}]^{-1}
		\end{split}
		\ene
		 is Shur stable, i.e., all the eigenvalues of $A^*+B^*K$ strictly lies in the unit circle. From this perspective, the data-driven property ID is strictly more data-efficient than that of model-based methods as here the explicit model is not necessary for identifying stabilizability. 
		
		A follow-up question is how to ensure the Shur stability of the closed-loop matrix $[\boldsymbol{x}_1^{(1)},\boldsymbol{x}_1^{(2)}][\boldsymbol{x}_0^{(1)},\boldsymbol{x}_0^{(2)}]^{-1}$. If for instance $\boldsymbol{x}_1^{(1)}=[0.5,1]'$ and $\boldsymbol{x}_1^{(2)}=[0,2]'$,  the system in \eqref{linearsys} however cannot be stabilized via the gain $K$ in the form of \eqref{fbgain}.  By \citet[Theorem 8]{van2020data}, we even cannot conclude whether $(A^*,B^*)$ is stabilizable or not via only using the control-state inputs in \eqref{input}. 
		
		Then, one may wonder whether it is still possible to identify stabilizability by simply changing the values of the control-state inputs in \eqref{input}. Unfortunately, the results of this work provide a negative answer and we have to increase the number of exciting input data.  \qed
		
	\end{example}
	
	\begin{example}(Controllability) 
		We consider an unknown linear system with $n=1$ and $m=1$.  Let $ u_0^{(1)}=1$ and ${x}_0^{(1)}=0$. Obviously, its controllability depends on whether the state feedback  ${x}_1^{(1)}$ is  non-zero or not, though the explicit model cannot be identified.  \label{secondexample}
	\end{example}
	
	In this work we study the property ID problem for the unknown linear system \eqref{linearsys}  in a direct data-driven approach with the minimum input.
	
	\subsection{Input sectional data, and  sets of linear systems}
	The {\em trajectory} data in the form of $\{\boldsymbol{x}_0,\boldsymbol{u}_0,\boldsymbol{u}_1,\ldots,\boldsymbol{u}_k\}$ is usually used for the system ID \cite{ljung1999system}. While the explicit model is unknown, it is not easy to ensure the stability of the closed-loop system, and thus is dangerous to collect a relatively long trajectory. Instead, we introduce the concept of  {\em input sectional data}, which consists of a series of (initial) state-control inputs of the form
\bee 
	X_-= \left[\boldsymbol{x}_0^{(1)},\boldsymbol{x}_0^{(2)},\ldots,\boldsymbol{x}_0^{(k)}\right],
U_-= \left[ \boldsymbol{u}_0^{(1)},\boldsymbol{u}_0^{(2)},\ldots,\boldsymbol{u}_0^{(k)}\right].\label{sectiondata}
\ene
	Note that only a pair of state-control input $(\boldsymbol{x}_0^{(i)},\boldsymbol{u}_0^{(i)})$ will be applied to the system per excitation. Though it is not trivial to freely set the initial state for any unknown system,  we can indeed achieve it in many cases, e.g., an inverted pendulum system. 	
	
	Let $\Sigma$ be the set of linear systems in the form of \eqref{linearsys}, i.e., 
	$$
	\Sigma=\{(A,B)|A\in\bR^{n\times n}, B\in\bR^{n\times m}\}.
	$$
	Given any  property $\mP$ of interest, we define $\Sigma_\mP\subseteq \Sigma$ as the set of linear systems with $\mP$. Clearly, a linear system has the property $\mP$ is equivalent to that it  belongs to $\Sigma_\mP$.  
	Since we are interested in the nontrivial property $\mP$, we always assume that $\Sigma_\mP$ is a non-empty proper subset of $\Sigma$ in this work and furthermore, $\Sigma_\mP$ can be either explicitly or implicitly  described as prior \citep{trentelman2012control}. For example, if $\mP$ denotes the stabilizability of linear systems in $\Sigma$, then 
	$$\Sigma_\mP=\{(A,B)\in \Sigma|\exists K\in\bR^{m\times n}~\text{s.t.}~\rho(A+BK)<1\}.$$
	 Let $\Sigma_\mP^c$ denote the complement set of $\Sigma_\mP$, i.e., the set of linear systems  in $\Sigma$  without property $\mP$. Thus, our objective reduces to check whether $(A^*,B^*)\in \Sigma_\mP$ or $(A^*,B^*)\in \Sigma_\mP^c$ via input sectional data in \eqref{sectiondata}. To this purpose, let the state {\em feedback}  of the system $(A^*,B^*)$ be
	\bee X_+ := [\boldsymbol{x}_1^{(1)},\boldsymbol{x}_1^{(2)},\ldots,\boldsymbol{x}_1^{(k)}]=A^*X_-+B^*U_-. \label{feedback}\ene
	Note that  $X_+$ cannot be implied from $(X_-,U_-)$ as $(A^*,B^*)$ is unknown.  Define the input-feedback sectional data  as
	\bee
	\mD=(X_-,U_-,X_+)
	\label{data}
	\ene
	and $\Sigma_\mD\subseteq \Sigma$ as the subset of linear systems obeying the dataset $\mD$, i.e., 
	\bee
	\Sigma_\mD =\left\{(A,B)\in\Sigma|X_+=AX_-+BU_-\right\}. \label{dataset}
	\ene
	Clearly, $(A^*,B^*)\in \Sigma_\mD$, and if  $\Sigma_\mD\subseteq \Sigma_\mP$ or $\Sigma_\mD\subseteq \Sigma_\mP^c$,  one can easily conclude whether $(A^*,B^*)$ has the property $\mP$ or not.  	
	However, if $\Sigma_\mD$ is not entirely contained in $\Sigma_\mP$ or $\Sigma_\mP^c$, it is unclear whether the system \eqref{linearsys} has the property $\mP$, the main reason of which is that the input sectional data $(X_-,U_-)$ is not {\em sufficiently rich} (c.f. Definition \ref{def_sr}) for identifying the property $\mP$. 	
	
	\subsection{Sufficient richness of input sectional data}
	Informally, we say the input sectional data $(X_-,U_-)$ is sufficiently rich for identifying a property $\mP$ if we can use it to excite the system  and conclude whether it has the property $\mP$ or not. To formalize it, 
	%
	define the input-feedback sectional data of a linear system $(A,B)\in\Sigma$ as
	\bee\label{datasystem}
	\mD(X_-,U_-|A,B)=(X_-,U_-,AX_-+BU_-),
	\ene
	which contains the input sectional data and the state feedback from the linear system $(A,B)$. 
	
	\begin{defi}
		The input sectional data $(X_-,U_-)$ is sufficiently rich for identifying a property $\mP$ of the unknown linear system \eqref{linearsys}  if either $\Sigma_{\mD(X_-,U_-|A,B)}\subseteq\Sigma_\mP$ or  $\Sigma_{\mD(X_-,U_-|A,B)}\subseteq\Sigma_{\mP}^{c}$ holds for any $(A,B)\in\Sigma$. \label{def_sr}
	\end{defi}
	
	We use the well-known identifiability property for illustration, the objective of which is to check whether the input sectional data $(X_-,U_-)$ is sufficiently rich for identifying the explicit  model $(A^*,B^*)$. Since 
	\bee\label{identfy}
	A^*X_-+B^*U_-=[A^*, B^*] \begin{bmatrix}X_-\\ U_-\end{bmatrix}=X_+,
	\ene
	the necessary and sufficient condition for the sufficient richness of $(X_-,U_-)$ for identifiability is 	
	\bee
	\text{rank}
[		X_-',U_-']'=n+m. 
	\label{fullrank}
	\ene
	For brevity, the above rank condition is said to be {\em persistently exciting} (PE) in the sequel.  For Example \ref{secondexample}, the sufficient richness condition for controllability is $U_-\neq 0$.  
	
	\subsection{The minimum input design of this work}
	Our objective  is to establish necessary and sufficient conditions for sufficient richness of the input sectional data $(X_-,U_-)$ for any property $\mP$, and show how to achieve sufficient richness with the {\em minimum} number of inputs, i.e., the minimum $k$ in \eqref{sectiondata}.  In comparison with \eqref{fullrank}, we can easily quantify the data-efficiency of the data-driven property ID over model-based methods. 
	
	Before proceeding, we show in Lemma \ref{half} that Definition \ref{def_sr} can be simplified. 
	\begin{lemma}
		The input sectional data $(X_-,U_-)$ is sufficiently rich for identifying a property  $\mP$ of the unknown linear system \eqref{linearsys}  if and only if  $\Sigma_{\mD(X_-,U_-|A,B)}\subseteq\Sigma_\mP$ for any $(A,B) \in \Sigma_\mP$. \label{half}
	\end{lemma}
	
	\begin{pf}($\Rightarrow$) Given any $(A,B)\in \Sigma_{\mP}$, it is trivial that $(A,B)\in\Sigma_{\mD(X_-,U_-|A,B)}$.  By Definition \ref{def_sr}, we can easily conclude that $\Sigma_{\mD(X_-,U_-|A,B)}\subseteq\Sigma_\mP$.
		
		($\Leftarrow$)  If $(X_-,U_-)$ is not sufficiently rich for identifying $\mP$, it follows from Definition \ref{def_sr} that there exists a linear system $(A_0,B_0) \in \Sigma_\mP^c$ such that $\Sigma_{\mD(X_-,U_-|A_0,B_0)}$ is not entirely contained in $\Sigma_\mP$ or  $\Sigma_{\mP}^{c}$.  That is, there must be some other linear system $(A_1,B_1)\in\Sigma_{\mD(X_-,U_-|A_0,B_0)}$ such that $(A_1,B_1) \in \Sigma_\mP$, which implies that 
		\bee \Sigma_{\mD(X_-,U_-|A_1,B_1)}\subseteq\Sigma_\mP.\label{systemset}
		\ene 
		Since both $(A_0,B_0)$ and $(A_1,B_1)$ are from the same set $\Sigma_{\mD(X_-,U_-|A_0,B_0)}$, it follows from \eqref{datasystem} that $${\mD(X_-,U_-|A_0,B_0)}={\mD(X_-,U_-|A_1,B_1)}.$$ Together with \eqref{systemset}, it holds that $\Sigma_{\mD(X_-,U_-|A_0,B_0)}\subseteq\Sigma_\mP$, which is a contradiction.
	\end{pf}
	
	\section{Design of minimum input sectional data for property ID}\label{sec_mini}
	
	For a matrix $M\in \bR^{m\times n}$, let $\text{im}(M)$ denote its image, i.e., $\text{im}(M)=\{Mx | \forall x\in\bR^{n}\}$. Given any input sectional data $(X_-,U_-)$, clearly $\text{im}([X_-', U_-']')$ is a linear subspace of $\bR^{n+m}$ and can be generated by a minimum set of linearly independent vectors, which is called a basis in linear algebra. Conversely, we can extract an input sectional data from any basis of a linear subspace of $\bR^{n+m}$ via the partition in conformity with $[X_-', U_-']'$.  
	 
	In this section, we shall use the linear subspace to characterize the necessary and sufficient condition for sufficient richness of an input sectional data, and use its basis to design the minimum input sectional data.  
	\subsection{The sufficient richness via the property-dependent minimum linear subspace}
	Our major result shows the existence of a property-dependent minimum linear subspace to check the sufficient richness of an input sectional data. 
	\begin{theo}\label{linearspace}
		For any property $\mP$ of the unknown linear system in \eqref{linearsys}, define the set of linear subspaces as 
		$$\hspace{-0.5cm}\Pi_\mP=\left\{\text{im}
		\begin{bmatrix}
			X \\ U
		\end{bmatrix}|(X,U)~\text{is sufficiently rich for identifying}~\mP\right\} $$
		and the intersection of all elements in $\Pi_\mP$ as 
		\bee
		\mL_\mP = \bigcap\nolimits_{\mL \in \Pi_\mP}\mL. \label{deflp}
		\ene
		Then an input sectional data $(X_-,U_-)$ is sufficiently rich for identifying $\mP$  {\em if and only if}
		\bee	\label{nsc}
		\text{im}\begin{bmatrix} X_- \\ U_-\end{bmatrix}\supseteq\mL_\mP.
		\ene 
	\end{theo}	
	The proof is given in the next subsection. Theorem \ref{linearspace} provides a necessary and sufficient condition for sufficient richness of an input sectional data via the linear subspace $\mL_\mP$ in \eqref{deflp}. In fact, $\mL_\mP$ is unique and contains the ``minimum" excitation input information required for identifying $\mP$. By Theorem \ref{linearspace}, it is trivial to establish the following results, the proofs of which are omitted. 
	\begin{coro}
		The minimum number $k$ of an input sectional data \eqref{sectiondata} for identifying property $\mP$ of the unknown linear system \eqref{linearsys} is $\text{dim}(\mL_\mP)$ where $\mL_\mP$ is given in \eqref{deflp}.
	\end{coro}
	\begin{coro}\label{coro2}
		Suppose that $[X_-',U_-']'$ is a basis of $\mL_\mP$, then $(X_-,U_-)$ is a minimum input sectional data for identifying property $\mP$ of the unknown linear system \eqref{linearsys}. 
	\end{coro}
	
	As in the system ID \cite{ljung1999system}, our results are established for {\em any} admissible feedback state $X_+$ in \eqref{feedback}. If $X_+$ is ``good" enough, it is possible (cf. Example \ref{firstexample}) that we can still identify $\mP$, even if the input section data $(X_-,U_-)$ does {\em not} satisfy  \eqref{nsc}.  Note that the value of $X_+$  is not known as a prior, and cannot be designed.  Sufficient richness condition in \eqref{nsc} can achieve property ID regardless of the value of $X_+$.
	
	If $\mL_\mP=\bR^{n+m}$, it follows from \eqref{fullrank} that an explicit model can be identified.  Thus, the explicit model is essentially needed for identifying $\mP$. Otherwise, the data-driven property ID is strictly more data-efficient than the model-based analysis. 
	
	\subsection{Proof of Theorem \ref{linearspace}}
	By \eqref{deflp} and \eqref{nsc}, we only need to prove the sufficiency of Theorem \ref{linearspace} and use the following lemmas.   	
	\begin{lemma}
		Consider a linear subspace $\mL\subseteq\bR^{n+m}$ that contains a sufficiently rich basis for identifying $\mP$ of the unknown linear system in \eqref{linearsys}, then
		\begin{enumerate}[(a)]
			\item any basis of $\mL$ is sufficiently rich for identifying $\mP$.
			\item an input sectional data  $(X_-,U_-)$ is sufficiently rich for identifying $\mP$ if
			\bee
			\text{im}\begin{bmatrix} X_- \\ U_-\end{bmatrix}\supseteq\mL.\label{containspace}
			\ene 
		\end{enumerate} 
		\label{sc}
	\end{lemma}
	\begin{pf} Suppose that $[\widehat{X}_-',\widehat{U}_-']'$ is a sufficiently rich basis of $\mL$  for identifying $\mP$.  
		
		{(a)}~Let $[\overline{X}_-',\overline{U}_-']'$ be a different basis of $\mL$.  Obviously, we have
		\bee
		\text{im}\begin{bmatrix} \overline{X}_- \\ \overline{U}_-\end{bmatrix} = \text{im}\begin{bmatrix}\widehat{X}_- \\ \widehat{U}_-\end{bmatrix}. \label{samespace}
		\ene
		For any $(\overline{A},\overline{B})\in\Sigma$, it follows from \eqref{dataset} and  \eqref{datasystem} that
		\bee
		\begin{split}
			&\Sigma_{\mD(\widehat{X}_-,\widehat{U}_-|\overline{A},\overline{B})} \\
			&= 
			\left\{(A,B) {\big |} [A,B]\begin{bmatrix}\widehat{X}_- \\ \widehat{U}_-\end{bmatrix}=[ \overline{A},\overline{B}]\begin{bmatrix}\widehat{X}_- \\ \widehat{U}_-\end{bmatrix}\right \} \\
			&=\left\{(A,B) {\big |} [A-\overline{A}, B-\overline{B}] \begin{bmatrix}\widehat{X}_- \\ \widehat{U}_-\end{bmatrix}=0 \right\}.
		\end{split}	\label{sd}
		\ene
		Jointly with \eqref{samespace},  it implies that
	$\Sigma_{\mD(\widehat{X}_-,\widehat{U}_-|\overline{A},\overline{B})} = \Sigma_{\mD(\overline{X}_-,\overline{U}_-|\overline{A},\overline{B})}.$ 
	By Definition \ref{def_sr}, then $(\overline{X}_-,\overline{U}_-)$ is also sufficiently rich for identifying $\mP$.
		
		{(b)} Since $(X_-,U_-)$ satisfies \eqref{containspace}, it follows from \eqref{sd} that $\Sigma_{\mD(X_-,U_-|\overline{A},\overline{B})} \subseteq \Sigma_{\mD(\widehat{X}_-,\widehat{U}_-|\overline{A},\overline{B})}$. As  $(\widehat{X}_-,\widehat{U}_-)$ is sufficiently rich for identifying  $\mP$, it follows from  Definition \ref{def_sr}  that $(X_-,U_-)$ is also sufficiently rich for $\mP$. 
	\end{pf}
	By Lemma \ref{sc}(a), there is no need to specify which basis of the linear subspace is used. Thus, we directly state that a linear subspace is sufficiently rich for identifying  $\mP$ if it contains a  basis that is sufficiently rich for identifying  $\mP$. 
	
	
	\begin{lemma} \label{pruning}
		Consider two subspaces $\mL_1$ and $\mL_2$  of $\bR^{n+m}$. If either subspace is sufficiently rich for identifying $\mP$ of the unknown linear system \eqref{linearsys}, then $\mL_1\cap\mL_2$ is also sufficiently rich for identifying $\mP$. 
	\end{lemma}
	\begin{pf} See Appendix \ref{sec:proofLemma3}. 
	\end{pf}
{\em Proof of Theorem \ref{linearspace}:}~Let $\mL_m\in\Pi_\mP$ be a linear subspace with the minimum dimension in $\Pi_\mP$. Then, we show that $\mL_{m}=\cL_\mP$. Clearly, it follows from \eqref{deflp} that $\mL_{m} \supseteq \cL_\mP$. For any subspace $\mL' \in \Pi_\mP$, we note from Lemma \ref{pruning} that $\mL_m\cap\mL'$ is also in $\Pi_\mP$. Since $(\mL_m\cap\mL')\subseteq\mL_m$ and $\mL_m$ is the subspace with the minimum dimension, then $(\mL_m\cap\mL')=\mL_m$, i.e., $v\in \mL' $ for any $v\in\cL_m$.  Since $\cL_\mP=\left\{v|v\in\cL',\ \forall\cL'\in\Pi_\mP\right\}$, it holds that $v\in\cL_\mP$   for any $v\in\cL_m$, i.e., $\mL_{m} \subseteq \cL_\mP$.  Jointly with Lemma \ref{sc}, the proof is completed.
	\subsection{Constructing $\mL_\mP$ from any sufficiently rich linear subspace}
	\label{sec_construct}
	%
	If we have a sufficiently rich subspace $\cL$, e.g., a trivial one is $\cL=\bR^{n+m}$, for identifying the property $\mP$ of the unknown linear system in \eqref{linearsys}, we can sequentially remove some basis vector(s) of $\cL$ to obtain $\cL_{\mP}$.  Particularly, $\cL=\cL_{\mP}$ if (a) $\cL$ is sufficiently rich, and (b) any proper subset of $\cL$ is not sufficiently rich. Note that the sufficient richness of $\cL$ can be checked by examining only one of its bases.  We shall adopt this principle to explicitly describe $\cL_{\mP}$ in the next section. 
	
	\section{Explicit minimum linear subspaces for property identification}\label{sec_sr}
	
	In this section, we first consider some  fundamental properties of linear systems, and explicitly describe their minimum subspaces. Then, we study the case of other structure ID problems. 
	
	\subsection{An explicit model is essentially needed for checking stabilizability and controllability}\label{sec_sc}
	In the data-driven control literature, e.g., \citet{de2019formulas,coulson2019data,berberich2020data,fazel2018global,zhao2021global}, they directly assume that the unknown linear systems are either stabilizable or controllable. Moreover, a  stabilizing gain is given as prior in \citet{fazel2018global}, which in fact is a nontrivial condition \citep{hu2022sample,perdomo2021stabilizing}. Moreover, we show that an explicit model is essentially needed to obtain a stabilizing gain. Particularly, checking stabilizability or controllability is almost as difficult as identifying the unknown system model.
	
	\begin{theo}[Stabilizability]
		Let $\mP$ denote stabilizability of the unknown  linear system in \eqref{linearsys}, then the associated minimum linear subspace in \eqref{deflp}  is $\mL_\mP=\bR^{n+m}$. \label{stabilizability}
	\end{theo}
	
	\begin{pf}
		Obviously, any basis of $\bR^{n+m}$ satisfies \eqref{fullrank}, and thus is sufficiently rich for identifying the explicit model, based on which stabilizability can be easily determined. 
		
		Let $\mL$ be any proper subspace of $\bR^{n+m}$, and assume that the input sectional data $(X_-, U_-)$ forms a basis of $\mL$. Then there exists a non-zero vector $\boldsymbol{h} \in \bR^{n+m}$ such that 
		\bee\boldsymbol{h}' [ X_-', U_-' ]'= \boldsymbol{0}.\label{nonzeroh}
		\ene
Let $\boldsymbol{h}'=[\boldsymbol{h}_{1:n}', \boldsymbol{h}_{n+1:n+m}']$ be partitioned into two blocks with the specified dimensions. We consider two cases separately. 

		{\bf Case I}: $\boldsymbol{h}_{1:n}={\bf 0}$. Then, $\boldsymbol{h}_{n+1:n+m}\neq {\bf 0}$.  Let  $$\overline{A}= \begin{bmatrix}
			{\bf e}_1' \\ \vdots 
		\end{bmatrix}
		\in \bR^{n\times n}, \overline{B}=
		\begin{bmatrix}
			\boldsymbol{h}'_{n+1:n+m} \\ 
			\vdots
		\end{bmatrix}
		\in \bR^{n\times m}$$
		where $\boldsymbol{e}_i\in\bR^{n}$ is the standard unit vector with $1$ on the $i$-th element, and  the unspecified elements are  zeros. 
		It is easy to check that
		$$
	[
		\overline{A}-\lambda I, \overline{B}] = 
	\left[\begin{array}{c|c|c}1-\lambda &  & \boldsymbol{h}_{n+1:n+m}' \\ \hline
		 &-\lambda \cdot I_{n-1}&  \end{array}
		\right]
		$$
		where the unspecified elements are also zeros. Since $\boldsymbol{h}_{n+1:n+m}\neq {\bf 0}$, it implies
		$\text{rank}[
		\overline{A}-\lambda I, \overline{B}]= n,\ \forall \lambda \in \mathbb{C}\ \text{with}\  |\lambda| \ge 1$.  Then, it follows from  the Hautus test  \citep{trentelman2012control} that  the system $(\overline{A},\overline{B})$ is stabilizable, e.g., $(\overline{A},\overline{B})\in\Sigma_{\mP}$. However, if $\lambda=1$, then
 		\bee\label{cs}
		\begin{split}
		\text{rank}(X_+- \lambda X_-)& = \text{rank}
		\left([
		\overline{A}- \lambda I, \overline{B}]
		\begin{bmatrix}
			X_- \\ U_-
		\end{bmatrix}\right)\\
		&=  \text{rank}
		\left(\begin{bmatrix}\boldsymbol{h}'\\  *
		\end{bmatrix}
		\begin{bmatrix}
			X_- \\ U_-
		\end{bmatrix}\right)\\
		&\leq n-1 		\end{split}
		\ene
		where $*$ denotes unspecified elements and the inequality follows from \eqref{nonzeroh}. By \citet[Theorem 8]{van2020data}, it implies that $\Sigma_{\mD(X_-,U_-|\overline{A},\overline{B})} \not\subseteq \Sigma_\mP$. In view of Lemma \ref{half}, we can obtain that $(X_-,U_-)$ is not sufficiently rich for identifying stabilizability.

		{\bf Case II}:  $\boldsymbol{h}_{1:n}\neq\boldsymbol{0}$, and assume that its first nonzero element is $h_{l}\neq 0, 1\le l\le n$. Consider the following system
		$$\overline{A}= \begin{bmatrix} \vdots\\
			-\frac{1}{h_l} \boldsymbol{h}_{1:n}'+ \boldsymbol{e}_l'\\  \vdots 		\end{bmatrix},~\text{and}~\overline{B}=
		\begin{bmatrix} \vdots\\ 
			-\frac{1}{h_l} \boldsymbol{h}'_{n+1:n+m} \\ 
			\vdots  
		\end{bmatrix},$$
		where the $l$-th rows in both matrices are the only nonzero rows. Obviously all the eigenvalues of $\overline{A}$ are zero, which implies that $(\overline{A},\overline{B})$ is stabilizable. Similar to {\bf Case I}, one can obtain that 
		$$
		\text{rank}(X_+- X_-) =  \text{rank}
		\left(\begin{bmatrix} * \\ -\frac{1}{h_l}\boldsymbol{h}'\\  *
		\end{bmatrix}
		\begin{bmatrix}
			X_- \\ U_-
		\end{bmatrix}\right)\leq n-1
		$$
		which implies that $(X_-,U_-)$ is not sufficiently rich for identifying stabilizability.
		
		Overall, any proper subspace $\mL$ is not sufficient rich for identifying stabilizability. By Section \ref{sec_construct}, we obtain $\mL_\mP = \bR^{n+m}$.
	\end{pf}
	
	\begin{theo}[Controllability]\label{controllability}
		Let $\mP$ denote controllability of the unknown linear system in \eqref{linearsys}. The associated minimum linear subspace in \eqref{deflp} is explicitly given as follows. 
		\begin{enumerate} [(a)]
			\item If $n=1$, then $\mL_\mP=\text{span}(\boldsymbol{e}_2,\boldsymbol{e}_3,...,\boldsymbol{e}_{m+1})$, where $\boldsymbol{e}_i\in\bR^{n+m}$ is the standard unit vector with $1$ on the $i$-th element and  $0$, otherwise.
			\item If $n>1$, then $\mL_\mP=\bR^{n+m}$. 
		\end{enumerate} 
	\end{theo}
	\begin{pf} 
		(a) For the case $n=1$, it follows from the Hautus test  \citep{trentelman2012control}  that \bee\label{sigmap}
		\Sigma_\mP = \left\{(A,B)\in \Sigma |B\neq\boldsymbol{0}'\right\}.
		\ene
		Let $(X_-,U_-)$ be formed by a basis of $\text{span}(\boldsymbol{e}_2,...,\boldsymbol{e}_{m+1})$, i.e., $X_-={\bf 0}'$ and $U_-\in\bR^{m\times m}$ is an invertible matrix. Let $(A_0,B_0)\in \Sigma_\mP$ and for any $(A_1,B_1)\in\Sigma_{\mD(X_-,U_-|A_0,B_0)}$, it follows from \eqref{dataset} and \eqref{datasystem} that $A_0X_-+B_0U_-=A_1X_-+B_1U_-$. This implies that $(B_0-B_1)U_-=\boldsymbol{0}'$, i.e., $B_1=B_0\neq\boldsymbol{0}'$. By \eqref{sigmap}, it is clear that $(A_1,B_1)\in \Sigma_\mP$ and $\Sigma_{\mD(X_-,U_-|A_0,B_0)}\subseteq \Sigma_\mP$. Then, it follows from Lemma \ref{half} that $(X_-,U_-)$ is sufficiently rich for identifying controllability.
		
		Next, we show that $\text{span}(\boldsymbol{e}_2,\boldsymbol{e}_3,...,\boldsymbol{e}_{m+1})$ is the minimum linear subspace. To this end, let $\boldsymbol{v}_1,...,\boldsymbol{v}_m \in \mathbb{R}^{m+1}$ be one of its bases, and  without loss of generality, we assume that $\boldsymbol{v}_m$ is dropped. Then there must exist a non-zero vector $\boldsymbol{h}\in\bR^{m+1}$ satisfying that 
		\bee\boldsymbol{h}_{2:m+1}\neq {\bf 0}~\text{and}~ \boldsymbol{h}'\boldsymbol{v}_j=0,\ j\in\{1,...,m-1\}.\label{hvector}
		\ene 
		Let $[\overline{A},\overline{B}] = \boldsymbol{h}'$, then $\overline{B}=\boldsymbol{h}_{2:m+1}'\neq 0$ and by \eqref{sigmap}, $(\overline{A},\overline{B})\in \Sigma_\mP$.  Consider the linear system $(A_0,B_0)$ with $A_0=0$ and $B_0=\boldsymbol{0}'$. It follows from \eqref{hvector} that ${\bf 0}=[\overline{A},\overline{B}] [\boldsymbol{v}_1,...,\boldsymbol{v}_{m-1}]=[\overline{A}_0,\overline{B}_0] [\boldsymbol{v}_1,...,\boldsymbol{v}_{m-1}]$. This implies that $(\overline{A}_0,\overline{B}_0)\in\Sigma_{\mD(\boldsymbol{v}_1,...,\boldsymbol{v}_{m-1}|\overline{A},\overline{B})}$. Obviously, $(\overline{A}_0,\overline{B}_0)  \notin\Sigma_\mP$. By Lemma \ref{half}, the linear subspace  $\text{span}(\boldsymbol{v}_1,...,\boldsymbol{v}_{m-1})$ is not sufficiently rich for controllability. In view of Section \ref{sec_construct}, we obtain $\mL_\mP=\text{span}(\boldsymbol{e}_2,\boldsymbol{e}_3,...,\boldsymbol{e}_{m+1})$.
		
		(b) For the case $n>1$,  let $\mL$ be a proper subspace of $\bR^{n+m}$, and assume that the input sectional data $(X_-, U_-)$ forms a basis of $\mL$. Similarly, there exists a non-zero vector $\boldsymbol{h} \in \bR^{n+m}$ such that 
		\bee\boldsymbol{h}' [ X_-', U_-' ]'= \boldsymbol{0}.\label{nonzerohc}
		\ene
		and we also consider two cases separately. 
		
		{\bf Case I}: $\boldsymbol{h}_{1:n}=\boldsymbol{0}$. Then, $\boldsymbol{h}_{n+1:n+m}\neq {\bf 0}$. Let
		\bee\label{caseiab}
		\overline{A}= \begin{bmatrix}
			1 & 0 & \cdots & 0 \\ 
			0 & 2 & \cdots & 0 \\ 
			\vdots & \vdots & \ddots & \vdots \\ 
			0 & 0 & \cdots & n 
		\end{bmatrix}, ~\text{and}~ \overline{B}=
		\begin{bmatrix}
			\boldsymbol{h}'_{n+1:n+m} \\ 
			\boldsymbol{1}' \\ 
			\vdots  \\ 
			\boldsymbol{1}' 
		\end{bmatrix}. 
		\ene
		If $\lambda \in \{ 1,...,n\}$, it is easy to check that $\text{rank}(\overline{A}-\lambda I) =n-1$. Jointly with that every row of $\overline{B}$ is nonzero, it implies that $\text{rank}[\overline{A}-\lambda I, \overline{B}]=n$. If $\lambda\notin \{ 1,...,n\}$, then $\text{rank}(\overline{A}-\lambda I) =n$, which trivially implies $\text{rank}[\overline{A}-\lambda I, \overline{B}]=n$. By the Hautus test, the system $(\overline{A},\overline{B})$ is controllable.

		{\bf Case II}: $\boldsymbol{h}_{1:n}\neq\boldsymbol{0}$, and there is no loss of generality to assume that $h_2\neq0$. Moreover, if $h_1=0$, let
		\bee\label{h1zero}
		\begin{split}
			\overline{A}= \begin{bmatrix}
				1 & 0 & 0 & \cdots & 0 \\ 
				0 & 1 & 0 & \cdots & 0 \\ 
				0 & 0 & 2 & \cdots & 0 \\ 
				\vdots & \vdots & \vdots& \ddots & \vdots \\ 
				0 & 0 & 0 & \cdots & n-1
			\end{bmatrix} +
			\begin{bmatrix}
				\boldsymbol{h}_{1:n}' \\ \boldsymbol{0}' \\ \boldsymbol{0}' \\ \vdots \\ \boldsymbol{0}'
			\end{bmatrix},
			\overline{B}=
			\begin{bmatrix}
				\boldsymbol{h}'_{n+1:n+m} \\ 
				\boldsymbol{1}' \\ 
				\vdots  \\ 
				\boldsymbol{1}'
			\end{bmatrix}.
		\end{split}
		\ene
		If $\lambda = 1$,  the second row is the only zero row in $\overline{A}-\lambda I$ and by $h_2\neq 0$, it is easy to check that $\text{rank}(\overline{A}-\lambda I)=n-1$. As the second row of $\overline{B}$ is nonzero,  we further obtain that $\text{rank}[\overline{A}-\lambda I, \overline{B}]=n$. If $\lambda\neq 1$, then $\overline{A}-\lambda I$ contains at most one zero row, which cannot be in the first two rows. While the corresponding row of $\overline{B}$ is nonzero, this implies that $\text{rank}[\overline{A}-\lambda I, \overline{B}]=n$.
		
		If $h_1\neq0$, let
		\bee\label{h1nozero}
		\overline{A}= \begin{bmatrix}
			1 & 0 & \cdots & 0 \\ 
			0 & 2 & \cdots & 0 \\ 
			\vdots & \vdots & \ddots & \vdots \\ 
			0 & 0 & \cdots & n
		\end{bmatrix} +
		\begin{bmatrix}
			\frac{1}{h_1}\boldsymbol{h}_{1:n}' \\ \boldsymbol{0}' \\ \vdots \\ \boldsymbol{0}'
		\end{bmatrix},
		\overline{B}=
		\begin{bmatrix}
			\frac{1}{h_1}\boldsymbol{h}'_{n+1:n+m} \\ 
			\boldsymbol{1}' \\ 
			\vdots  \\ 
			\boldsymbol{1}' 
		\end{bmatrix}.
		\ene
		We can follow the same arguments as the case of  $h_1 = 0$ to conclude that 
$\text{rank}[(\overline{A}-I)-\lambda I, \overline{B}]=n$ for any $\lambda\in\mathbb{C}$. That is,  $\text{rank}[\overline{A}-\lambda I, \overline{B}]=n$  for any $\lambda\in\mathbb{C}$. Thus, we can use the Hautus test to confirm that the two systems in \eqref{h1zero} and \eqref{h1nozero} are controllable.

Similar to \eqref{cs}, one can also derive that $
			\text{rank}(X_+- X_-) \le n-1$ for any system in \eqref{caseiab}-\eqref{h1nozero}. By \citet[Theorem 8]{van2020data}, it implies $\Sigma_{\mD(X_-,U_-|\overline{A},\overline{B})} \not\subseteq \Sigma_\mP$. Jointly with Lemma \ref{half},  $(X_-,U_-)$ is not sufficiently rich for identifying controllability.

		Overall, any proper subspace $\mL$ of $\bR^{n+m}$ is not sufficiently rich for identifying controllability. By Section \ref{sec_construct}, we obtain $\mL_\mP = \bR^{n+m}$ for the case of $n>1$.
	\end{pf}
	
	Both Theorem \ref{stabilizability} and Theorem \ref{controllability} reveal that we essentially have to use $n+m$ inputs for identifying stabilizability and controllability, except for the trivial case that $n=1$. Note from Example \ref{firstexample} that we can still possible to identify stabilizability before using $n+m$ samples. This suggests that an online input design method could be more data-efficient.

%
%
%
%
	
	\subsection{Sufficient richness for structure identification}
	Though the results in Section \ref{sec_sc} appear to be disappointing, we show in this subsection that it is not the case for many structural properties of linear systems. 
	
Firstly, we aim to determine whether the unknown matrices $A$ and $B$ have some specified structure of interest, i.e.,  
	\bee \label{stru_sigma_p}
	\Sigma_\mP = \{(A,B)|a_{ji}=b_{pq}=0, \forall (j,i) \in \mathcal{I}_A, (p,q) \in \mathcal{I}_B\},
	\ene
	where $\mathcal{I}_A$ and $\mathcal{I}_B$ denote the positions of the zero elements of interest in $A$ and $B$, respectively. This also characterizes their sparsity pattern.  We can explicitly describe the minimum linear subspace $\mL_\mP$  in \eqref{deflp}.
	
	\begin{prop} \label{sr_stru}
		Consider the structural property of the unknown linear system \eqref{linearsys} that is specified in \eqref{stru_sigma_p}. Let \bee\mathcal{I}_\mP = \{i|\exists (j,i) \in \mathcal{I}_A \text{ or } (p,i-n) \in \mathcal{I}_B\}.\label{indexset}\ene
		That is, $\mathcal{I}_\mP$ denotes the indices of columns with the zero elements of interest in the augmented matrix $[A,B]$. Then the associated minimum linear subspace in \eqref{deflp}  is $\mL_\mP = \text{span}\{\boldsymbol{e}_i,\ i\in\mathcal{I}_{\mP}\}.$
	\end{prop}
	\begin{pf} See Appendix \ref{prop}.
	\end{pf}
	
Let $x_k$ and $u_k$ be the state and input vectors of a network with $n$ nodes. Then, we can use Proposition \ref{sr_stru} to examine whether there exist interactions among some specified pairs of nodes.  Suppose that we want to identify if the state of node $i$ will affect the state of node $j$, i.e., $a_{ji}\not=0$. Then, Proposition \ref{sr_stru} implies that $\mL_\mP = \text{span}\{\boldsymbol{e}_i\}$. That is, the input section data $(X_-,U_-)$ in \eqref{sectiondata} only contains one vector. 
	
Secondly, we extend \eqref{stru_sigma_p} to more general case whether the linear combinations of elements in $[A,B]$ have specified structural property.  To this purpose, we define a subset of linear systems with their parameters that are linearly constrained, i.e., 
	\bee \label{def_li}
	\Sigma_i = \left\{(A,B)\in\Sigma|\boldsymbol{h}_i'\text{vec}\left(\left[A, B\right]\right)\in\mathcal{S}_i \right \}, i\in\{1,\ldots,\ell\}
	\ene
	where $\boldsymbol{h}_i$ is a vector in $\bR^{n(n+m)}$, $\text{vec}\left(\cdot\right)$ denotes the vectorization operator to stack the columns of a matrix on top of one another, and $\mathcal{S}_i$ is a proper subset of $\bR$.  A broad class of structurally constrained linear systems of our interest contains the intersection of these subsets, e.g.,  $\Sigma_\mP=\cap_{i=1}^\ell \Sigma_i$, which covers all the polyhedral constraints on $\text{vec}\left(\left[A, B\right]\right)$. Note that \eqref{stru_sigma_p} is also included in this form.

	\begin{theo}\label{general} Suppose that the set of linear systems
	\bee
	\Sigma_\mP=\cap_{i=1}^\ell \Sigma_i\label{intersection}
	\ene
	is not empty.  Then, the associated minimum linear subspace in \eqref{deflp} is $\mL_\mP=\text{im}(M)$ where 
		\bee \label{def_M}
		M = \begin{bmatrix}\text{vec}^{-1}(\boldsymbol{h}_1) \\ \text{vec}^{-1}(\boldsymbol{h}_2) \\ \vdots \\ \text{vec}^{-1}(\boldsymbol{h}_\ell)\end{bmatrix}'~\text{and}~\text{vec}^{-1}(\cdot)\in \mathbb{R}^{n\times(n+m)}
		\ene
	is the inverse of the vectorization operator $\text{vec}(\cdot)$.
	\end{theo}
	\begin{pf}
		Let
		$[X_-',U_-']'= M$ and $(A_0,B_0)\in \Sigma_\mP$. For any $(A_1,B_1) \in \Sigma_{\mD(X_-,U_-|A_0,B_0)}$, it follows from \eqref{dataset} and \eqref{datasystem} that  
		\begin{equation}
			[A_1, B_1]M = [A_0, B_0]M. \label{t51}
		\end{equation}
	We rewrite both sides of (\ref{t51}) in the form of block matrices, i.e., 
		\begin{equation}
			\left[C_1^{(1)},\ldots,C_1^{(\ell)}
			\right]=\left[C_0^{(1)}, \ldots, C_0^{(\ell)}\right]\label{t52}
		\end{equation}
		where $C_1^{(\cdot)}\in \bR^{n\times n}$.
		Then, $\text{tr}(C_1^{(i)})$ is just $\boldsymbol{h}_i^T\text{vec}\left(\left[A_1, B_1\right]\right)$ and (\ref{t52}) implies that
		\been
\boldsymbol{h}_i'\text{vec}\left(\left[A_1, B_1\right]\right) = \boldsymbol{h}_i'\text{vec}\left(\left[A_0, B_0\right]\right)\in\mathcal{S}_i,\forall i\in\{1,\ldots,\ell\}. 
		\enen
		That is, $\Sigma_{\mD(X_-,U_-|A_0,B_0)}\subseteq\Sigma_\mP$. By Lemma \ref{half}, the subspace $\text{im}(M)$ is sufficiently rich for identifying whether an unknown linear system of  $\Sigma$  belongs to $\Sigma_\mP$.

		Next, let $\mL$ be any proper subspace of $\text{im}(M)$ and $\boldsymbol{v}_1,...,\boldsymbol{v}_{p}$ be one of its bases.  Then, there must exist a column $\boldsymbol{w}$ of $M$ such that $\boldsymbol{w}\not\in\mL$, and we denote it as the $j$-th column of $\{\text{vec}^{-1}(\boldsymbol{h}_l)\}'$ for some  $l\in\{1,\ldots,\ell\}$. Define $\boldsymbol{h}=\boldsymbol{w}-\text{proj}_{\cL}(\boldsymbol{w})$ where $\text{proj}_{\cL}(\boldsymbol{w})$ projects the vector $\boldsymbol{w}$ to the subspace $\cL$. Since $\boldsymbol{w}\not\in\mL$, then $\boldsymbol{h}$ is a nonzero vector and orthogonal to $\cL$. Jointly with the fact that $\boldsymbol{w}\not\in\mL$, it holds that  
		\bee\boldsymbol{h}'\boldsymbol{v}_i=0,i\in\{1,\ldots,p\},~\text{and}~
		\boldsymbol{h}'\boldsymbol{w} \not=0.\label{hvector} 
		\ene
		 For any $c\in\bR$, let 
		\bee\label{barAB}
		\left[\overline{A}, \overline{B}\right] = \left[A_0, B_0\right] + \left[\ldots, c\boldsymbol{h},\ldots \right]',
		\ene
		where the second term of the right hand side (RHS) only has a nonzero  column in the $j$-th column. It follows from \eqref{hvector} that
		 $$\left[\overline{A}, \overline{B}\right]\boldsymbol{v}_i=\left[A_0, B_0\right]\boldsymbol{v}_i,\ i\in\{1,\ldots,p\}.$$
	That is,  $(\overline{A}, \overline{B})\in\Sigma_{\mD(\boldsymbol{v}_1,\ldots,\boldsymbol{v}_p|A_0,B_0)}$.  In view of \eqref{barAB}, it is trivial that
	 		\been
		\boldsymbol{h}_l'\text{vec}([\overline{A}, \overline{B}]) = \boldsymbol{h}_l'\text{vec}\left(\left[A_0,  B_0\right]\right) + c\cdot\boldsymbol{h}'\boldsymbol{w}.
		\enen
		Since $\mathcal{S}_l$ is a proper subset of $\bR$ and $ \boldsymbol{h}'\boldsymbol{w} \not= 0$ there must exist some $c\in\bR$ such that $\boldsymbol{h}_l'\text{vec}([\overline{A}, \overline{B}]) \notin \cS_l$. 
This implies that $(\overline{A}, \overline{B})\not\in\Sigma_\mP$. By Lemma \ref{half}, $\mL$ is not sufficiently rich for identifying $\mP$. In view of Section \ref{sec_construct}, it immediately holds that $\mL_\mP=\text{im}(M)$.	
	\end{pf}
	
	We use the following example to illustrate our result in Theorem \ref{general}. 
	\begin{example} \label{exam_for_t4}
		Consider a second-order autonomous system
		$$\boldsymbol{x}_{t+1} = \begin{bmatrix}
			a_{11} & a_{12} \\
			a_{21} & a_{22}
		\end{bmatrix} \boldsymbol{x}_t.$$
		To identify the structure that $a_{11} + a_{22}=0$, let $\boldsymbol{h}_1=[1,0, 0, 1]'$ and $\mathcal{S}_1=\{0\}$.  By Theorem \ref{general}, we  obtain that $M=I_2$ and $\mL_\mP = \bR^2$, which can also be used to identify the system parameters. However, it is not the case for the following structural properties, including 
		\begin{itemize}
			\item $\mP$: $a_{11} + a_{12} = 0$, $\mL_\mP=\text{im}\left[1,1\right]'$.
			\item $\mP$: $a_{11} + a_{21} = 0$, $\mL_\mP=\text{im}\left[1,0\right]'$.
			\item $\mP$: $a_{11} + a_{12} + a_{21} + a_{22} = 0$, $\mL_\mP=\text{im}\left[1,1\right]'$.
			\item $\mP$: $a_{11} + a_{12} = 0\ \text{and} \ a_{21} + a_{22} = 0$, $\mL_\mP=\text{im}\left[1,1\right]'$.
		\end{itemize}
		It is interesting to note from this example that the property on the sums of $a_{11}+a_{22}$ and $a_{12}+a_{21}$ are more difficult to identify than those of the other two elements. 
		\end{example}
	
Unfortunately,  the intersection operation in \eqref{intersection} excludes many structural properties, e.g., \{$a_{11}=0$ or $b_{11}=0$\}. We further extend to the following set of structurally constrained linear systems 
	\bee
	\Sigma_\mP=\Sigma_1\odot \Sigma_2\odot \ldots \odot \Sigma_\ell,  \odot\in\{\cap,\cup\}\label{interunion}
	\ene
	and obtain the following result. 
	
	\begin{theo} Suppose that  $\cS_i$ for any  $i\in\{1,\ldots,\ell\}$ in \eqref{def_li} is a bounded and non-empty set and $\{\boldsymbol{h}_1,\ldots,\boldsymbol{h}_\ell\}$ are linearly independent. Then the associated minimum linear subspace in \eqref{deflp} is $\mL_\mP=\text{im}(M)$ where $M$ is given in \eqref{def_M}.
		  		\label{general2}
	\end{theo}
	\begin{pf} See Appendix \ref{sec::proofTheo5}. 
	\end{pf} 
	
	\begin{rem} \label{brackets}
		Note that  \eqref{interunion} does not include some common cases, e.g., $\Sigma_\mP = (\Sigma_1 \cup \Sigma_2) \cap (\Sigma_3 \cup \Sigma_4)$ for $\ell=4$. In fact, the results in Theorem \ref{general2} still hold even we arbitrarily change the order of set operators in  \eqref{interunion}, e.g., $\Sigma_\mP=\left(\Sigma_1\odot\left( \Sigma_2\odot \ldots \right)\odot \Sigma_\ell \right),  \odot\in\{\cap,\cup\}$, where we use the brackets to change the order of intersection and union operators. The detailed proof of this case is very involved and we provide its sketch in Appendix \ref{app_brackets}.
		
	\end{rem}
\section{Direct data-driven property identification} \label{sec_dd}
	If $\cL_\cP\not= \mathbb{R}^{m+n}$, it follows from \eqref{fullrank} that one cannot explicitly identify the model parameters of \eqref{linearsys}. However, we can still use the {\em model-based approach} to achieve property ID. Specifically, we first excite the unknown system \eqref{linearsys} with a sufficiently rich input sectional data and then {\em arbitrarily} select a linear system from the set \eqref{dataset}.  By the sufficient richness of the input sectional data (cf. Definition \ref{def_sr}), the property ID result for both the unknown system \eqref{linearsys} and the selected system is of the same. Thus, we can complete the property ID task by using the linear system theory on the selected linear system. 	
	
	This section  shows how to complete the direct data-driven property ID for the unknown system \eqref{linearsys} with the sufficiently rich input sectional data. In \citet{van2020data}, the data enabled conditions for stabilizability and controllability have been established for the {\em set} of linear systems in  \eqref{dataset}. However, they cannot be used for the property ID of an individual linear system since their input data may not always be sufficiently rich. This work has resolved the input design problem, and under the excitation of the sufficiently rich input, their results can be directly used to identify stabilizability and controllability of the unknown linear system of \eqref{linearsys}. Thus,  we only consider the direct data-driven ID problem of the structure specified in \eqref{stru_sigma_p}. 
	
	\begin{theo} \label{info_stru}
		Let $\mP$ denote the structural property defined in \eqref{stru_sigma_p}. Suppose that the input sectional data $(X_-,U_-)$ in \eqref{sectiondata} is sufficiently rich for identifying $\mP$(c.f. Theorem \ref{linearspace} and Proposition \ref{sr_stru}) and $Q$ is selected to satisfy that \bee \label{def_Q}[X'_-,U'_-]'Q = [\boldsymbol{e}_{i},i\in\cI_\mP].\ene
		 Then the unknown system in \eqref{linearsys} has the property $\mP$ if and only if 
		$X_+Q\in \cM_{\mP}$, where $X_+$ is given in \eqref{feedback} and \bee\label{def_Mp}\cM_{\mP}=\{[A,B][\boldsymbol{e}_{i},i\in\cI_\mP]~|~ \forall (A,B)\in\Sigma_\mP\}.\ene
	\end{theo}
	\begin{pf}
	 	See Appendix \ref{append_id}.
	\end{pf}
Though  $Q$ in \eqref{def_Q} is not unique,  the result of Theorem \ref{info_stru} does not depend on its particular selection. 
	
	\begin{example} \label{example_stru}
		We aim to verify whether the unknown linear system with
				\bee\label{truesys} A^*=\begin{bmatrix}
			0 ~ 1\\2 ~ 1
		\end{bmatrix}~\text{and}~B^*=\begin{bmatrix}
			1 \\ 0
		\end{bmatrix}\ene
		belongs to the following set of linear systems 
		\bee\label{sigmapex}
	\Sigma_\mP = \{(A,B)|a_{11}=b_{12}=0\}.
	\ene
By Proposition \ref{sr_stru} and Theorem \ref{info_stru}, $\mL_\mP=\text{span}(\boldsymbol{e}_1, \boldsymbol{e}_3)\subseteq \bR^{3}$ and $\cM_\mP=\left\{M\in\bR^{2\times 2}| m_{11}=m_{22}=0\right\}$. 

Then, we choose an input sectional data 
		$X_-=\begin{bmatrix}
			1~0\\0~0
		\end{bmatrix}, U_-=[0~1].$
		It  follows from \eqref{linearsys} that $X_+=\begin{bmatrix}
			0~1\\2~0
		\end{bmatrix}$. 
		Since  $[X'_-,U'_-]'= [\boldsymbol{e}_1, \boldsymbol{e}_3]$, then $Q=I_2$ and $X_+Q\in \cM_\mP$.  By Theorem \ref{info_stru},  it implies that the system in \eqref{truesys} indeed belong to $\Sigma_\mP$ in \eqref{sigmapex}. 
		
	Consider another unknown system 
		 \bee\label{truesys2} A^*=\begin{bmatrix}
			1~ 1\\2 ~ 1
		\end{bmatrix}~\text{and}~B^*=\begin{bmatrix}
			1 \\ 0
		\end{bmatrix}.\ene 
Then, $X_+=\begin{bmatrix}
			1~1\\2~0
		\end{bmatrix} \notin \cM_\mP$ and we conclude that it does not belongs to $\Sigma_\mP$ in \eqref{sigmapex}. In both cases, we cannot identify the system parameters but immediately complete the property ID in a direct data-driven fashion. 
	\end{example}
	
	\section{Conclusion and future works} \label{sec_dc}
	Via a new concept of sufficient richness of input sectional data, we have explicitly shown how to design the minimum input data for identifying several important properties of an unknown linear system in a direct data-driven approach. It is encouraging to note its advantages over the model-based methods in the system analysis and design.  While this work only considers the noiseless system, it can be extended to noisy systems, e.g., the system \eqref{linearsys} has bounded process noises.

Moreover, we highlight some interesting observations that deserve further investigation. 
\begin{itemize}
\item	Usually, there are two settings for the input design. One is the {\em offline} setting, where we design the input sectional data $(X_-,U_-)$  without  using any state feedback $X_+$. The other is the {\em online} setting where the input sectional data is designed sequentially via the state feedback $X_+$.  One may wonder if the online feedback can reduce the sample complexity for the property ID. The answer is not conclusive as shown in the following example.
	
	\begin{example} \label{exam_online}
		Consider an unknown linear system with $n=2$ and $m=1$. We aim to examine the structure that $\{a_{11}=0,a_{12}=0\}$ or $\{a_{11}\neq0,b_1=0\}$. Similar to Theorem \ref{general2}, we can prove that the associated minimum linear subspace is $\mL_\mP=\bR^{3}$.   However, if we use $\boldsymbol{e}_1$ as the first input, we then are able to determine whether $a_{11}=0$. If $a_{11}=0$, then we select $\boldsymbol{e}_2$ as the next input, and $\boldsymbol{e}_3$ otherwise. In either case, we only need two state-control inputs for property ID. 	If the  structure of interest is $\{a_{11}=0, a_{12}=0 \text{ or } b_1=0\}$, then the online setting does not help to reduce the sample complexity. 
	\end{example}

\item If we have some prior information on the unknown linear system, e.g., $(A^*,B^*)$ is known from an  a priori  proper subset of $\Sigma$, then Theorem \ref{linearspace} may not hold. Take the scalar linear system as an example where the prior information is that there is just one zero on  $[A,B]$. Suppose we aim to identify the structure that $\{A=0\}$.   Clearly, either $(X_-,U_-)=(0,1)$ or $(X_-,U_-)=(1,0)$ is sufficiently rich for our property ID.  Since the two inputs form two different one-dimension linear subspaces and their intersection is a null space, Lemma \ref{pruning} does not hold any more.

\item Since the initial state may not always be reset for some unknown systems, we have to use the trajectory data in the form of $\{\boldsymbol{x}_0,\boldsymbol{u}_0,\boldsymbol{u}_1,\ldots,\boldsymbol{u}_k\}$. In this case, some properties of our interest cannot be identified. For an extreme case, let 
	$$
	\boldsymbol{x}_{t+1}=\begin{bmatrix}
		0.5&0\\0&0.5
	\end{bmatrix}\boldsymbol{x}_t+\begin{bmatrix}
		0\\0
	\end{bmatrix}u_t.
	$$ It seems that we can only conclude the uncontrollability of the system.  Then, the sufficient richness of this case is much more involved. 
	\end{itemize}
	
	\appendix
	
	\section{Proof of Lemma \ref{pruning}}
	\label{sec:proofLemma3}

%
%
%
%
%

	Suppose that $[X_-',U_-']$ is a basis of $\mL_1\cap\mL_2$.  Then,  we extend it to form the bases of $\mL_1$ and $\mL_2$ respectively, i.e., 
	\bee\label{abase}
	\begin{bmatrix}\overline{X}_- \\ \overline{U}_-\end{bmatrix}=
	\left[
	\begin{array}{c|c}  X_- & \overline{X}^{c}_- \\ U_- & \overline{U}^{c}_-\end{array}\right], ~\begin{bmatrix}\widetilde{X}_- \\ \widetilde{U}_-\end{bmatrix}= \left[
	\begin{array}{c|c}   X_- & \widetilde{X}^{c}_- \\ U_- & \widetilde{U}^{c}_-\end{array}\right].\ene
If $\mL_1\subseteq\mL_2$ or $\mL_2\subseteq\mL_1$, the result obviously holds. Thus, we assume that neither $\mL_1$ nor $\mL_2$ includes the other in the proof. Jointly with \eqref{abase}, it follows that
	$$\forall \begin{bmatrix} x \\ u\end{bmatrix} \in \text{im}\begin{bmatrix} \widetilde{X}^{c}_- \\  \widetilde{U}^{c}_-\end{bmatrix}, \begin{bmatrix} x \\ u\end{bmatrix} \not\in \text{im}\begin{bmatrix} \overline{X}_- \\  \overline{U}_-\end{bmatrix}.$$
	This implies that the matrix $\begin{bmatrix}
		\overline{X}_-&\widetilde{X}^{c}_-\\ \overline{U}_-&\widetilde{U}^{c}_-
	\end{bmatrix}$ has full column rank. Given any $(A,B)\in \Sigma_\mP$ and any $(A_1,B_1)\in\Sigma_{\mD(X_-,U_-|A,B)}$, it is clear that the following linear equation has a solution with respect to $[A_0,B_0]$,
		\been
			\begin{bmatrix}
				A_0,B_0
			\end{bmatrix}
			\begin{bmatrix}
				\overline{X}_-&\widetilde{X}^{c}_-\\ \overline{U}_-&\widetilde{U}^{c}_-
			\end{bmatrix} = \begin{bmatrix}
				A\overline{X}_- + B\overline{U}_-, A_1\widetilde{X}^{c}_- + B_1\widetilde{U}^{c}_-
			\end{bmatrix}.
	\enen
	Note that $(A_0,B_0)\in \Sigma_{\mD(\overline{X}_-,\overline{U}_-|A,B)}\subseteq\Sigma_\mP$. Jointly with $(A_1,B_1)\in\Sigma_{\mD(X_-,U_-|A,B)}$, i.e., $A_1X_-+ B_1U_- = AX_- + BU_-$, it implies that
	\been
			\begin{split}
			[A_1,B_1]
		\begin{bmatrix}
				\widetilde{X}_-\\ \widetilde{U}_-
			\end{bmatrix} 
			&= [AX_- + BU_-, A_0\widetilde{X}^{c}_- + B_0\widetilde{U}^{c}_-]\\
			&= [A_0X_- + B_0U_-, A_0\widetilde{X}^{c}_- + B_0\widetilde{U}^{c}_-] \\
			&= \begin{bmatrix}
				A_0,B_0
			\end{bmatrix} 
			\begin{bmatrix}
				\widetilde{X}_-\\ \widetilde{U}_-
			\end{bmatrix}.
		\end{split}
		\enen
Thus, $(A_1,B_1) \in \Sigma_{\mD(\widetilde{X}_-,\widetilde{U}_-|A_0,B_0)}$ and $\Sigma_{\mD(X_-,U_-|A,B)}\subseteq\Sigma_{\mD(\widetilde{X}_-,\widetilde{U}_-|A_0,B_0)}$.  Since $(\widetilde{X}_-,\widetilde{U}_-)$ is sufficiently rich and $(A_0,B_0)\in \Sigma_\mP$, we obtain that $\Sigma_{\mD(X_-,U_-|A,B)}\subseteq\Sigma_\mP$.  By Lemma \ref{half} we can conclude that $(X_-,U_-)$ is sufficiently rich for identifying $\mP$.
	
	\section{Proof of Proposition \ref{sr_stru}}
	\label{prop}
		We use  Theorem \ref{general} to complete the proof. Let $\mathcal{S}_r = \{0\}$, $r\in \{1,...,\ell\}$, where $\ell$ denotes the size of $\mathcal{I}_\mP$,  and $\{\boldsymbol{h}_1, \ldots, \boldsymbol{h}_{\ell}\}= \{\boldsymbol{e}_{(j-1)(m+n)+i},\boldsymbol{e}_{(p-1)(m+n)+n+q}, (j,i) \in \mathcal{I}_A, (p,q) \in \mathcal{I}_B\}$. Then, one can verify that $\Sigma_\mP$ in \eqref{intersection} is the same as \eqref{stru_sigma_p}.
		Note that the matrix $\text{vec}^{-1}(\boldsymbol{h_r})$ has only one nonzero entry for any $r\in\{1,...,\ell\}$, and its position is either $(j,i)\in \mathcal{I}_A$ or $(p,n+q)$ if $(p,q) \in \mathcal{I}_B$. Thus, the nonzero column vectors of $M$ in \eqref{def_M} are $\{\boldsymbol{e}_i,\ i\in\mathcal{I}_\mP\}.$ By Theorem \ref{general}, it follows that $\mL_\mP = \text{im}(M) = \text{span}\{\boldsymbol{e}_i,\ i\in\mathcal{I}_\mP\}.$ 
		
	\section{Proof of Theorem \ref{general2}}  \label{sec::proofTheo5}
	
	By Theorem \ref{general}, we can show that  $\text{im}(M)$ is sufficiently rich for identifying whether $(A^*,B^*) \in \Sigma_i$ for any $i\in\{1,...,\ell\}$ and thus for identifying $\mP$.
	
	To show the minimality of $\text{im}(M)$, let $\mL$ be any proper subspace of $\text{im}(M)$ and $\boldsymbol{v}_1,...,\boldsymbol{v}_{p}$ be one of its bases. 
		
	
	First, let $\boldsymbol{w}$  and $\boldsymbol{h}$ be given in the proof of Theorem  \ref{general}.  We partition the set $\{1,2,...,\ell\}$ into two disjoint subsets $\mathcal{C}_1$ and $\mathcal{C}_2$ where 
	$$\mathcal{C}_1= \{i|\text{vec}^{-1}(\boldsymbol{h}_i)\boldsymbol{h} \neq \boldsymbol{0},i=1,\ldots,\ell\}.$$
Since $\boldsymbol{h}'\boldsymbol{w} \not=0$ and $\boldsymbol{w}'$ is a row of $\text{vec}^{-1}(\boldsymbol{h}_l)$ for some $l\in\{1,\ldots,\ell\}$, it holds that $\text{vec}^{-1}(\boldsymbol{h}_l)\boldsymbol{h} \neq \boldsymbol{0}$ and $\cC_1$ is not empty. 
	
	Then, we choose a system $(A_0,B_0)$ from the set
	\bee \label{constraint_ab} (A_0,B_0)\in \cap_{i=1}^\ell \Sigma_i',\ene
	where $\Sigma_i'$ is either $\Sigma_i$ or $\Sigma_i^c$ and decided by Algorithm \ref{algorithm1}, and show that $(A_0,B_0)\in \Sigma_\mP$.  Note that $\{\boldsymbol{h}_1,\ldots, \boldsymbol{h}_\ell\}$ is linearly independent and $\cS_i$ is a bounded and non-empty set. It implies that $\cap_{i=1}^\ell \Sigma_i'$ is not an empty set. 
	
	To show $(A_0,B_0)\in \Sigma_\mP$, we note that Algorithm 1 terminates in either Line 8 or 15. If it terminates in Line 8 with $i=i_0$, one can verify that $(A_0,B_0)\in\Sigma_\mP$ if and only if 
	\bee \label{t55} (A_0,B_0) \in \Sigma_1 \odot \Sigma_2\odot \ldots \odot \Sigma_{i_0}, \ene
	where the operator of $\odot$ is the same as that in \eqref{interunion}. 
	\begin{algorithm}[t!]
			\caption{Construction of $\Sigma_i'$ for all $i\in\{1,\ldots,\ell\}$. }
			\label{algorithm1}
		\hspace*{0.02in} {\bf Input:}
		$\Sigma_\mP$ in \eqref{interunion}.\\
		\hspace*{0.02in} {\bf Output:} 
		$\Sigma_r',r=1,...,\ell.$
		\begin{algorithmic}[1]
			\For{$i=\ell$ to 1}
			\If{$i\in\mathcal{C}_1$ and $i\neq1$}
			\State let $\Sigma_i'=\Sigma_i$;
			\If{the operator before $\Sigma_i$ is $\cup$}
			\State let $\Sigma_j'=\Sigma_j^c$ for all $j\in\{1,\ldots,i-1\}$; 
			\Else
			\State let $\Sigma_j'=\Sigma_j$ for all $j\in\{1,\ldots,i-1\}$;
			\EndIf
			\State \textbf{break}
			\ElsIf{$i\in\mathcal{C}_2$}
			\If{the operator before $\Sigma_i$ is $\cup$}
			\State let $\Sigma_i'=\Sigma_i^c$; 
			\Else
			\State let $\Sigma_i'=\Sigma_i$; 
			\EndIf
			\Else
			\State let $\Sigma_1'=\Sigma_1$.
			\EndIf
			\EndFor
		\end{algorithmic}
	\end{algorithm}
	Moreover, if the operator before $\Sigma_{i_0}$ is $\cap$ in \eqref{interunion}, then Line 7 implies that $(A_0,B_0) \in \Sigma_1 \cap \Sigma_2 \cap \ldots \cap \Sigma_{i_0}$, which further implies \eqref{t55}. If the operator before $\Sigma_{i_0}$ is $\cup$, then $(A_0,B_0)\in \Sigma_{i_0}$ directly implies \eqref{t55}. In any case, we obtain that $(A_0,B_0)\in\Sigma_\mP$. 
	
If Algorithm \ref{algorithm1} terminates in Line 15, one can also show that $(A_0,B_0)\in\Sigma_\mP$ if and only if $(A_0,B_0) \in \Sigma_1$. By Line 15 we can also obtain that $(A_0,B_0)\in\Sigma_\mP$.

	Finally, we consider 
	\bee \label{barAB2}
	\left[\overline{A}, \overline{B}\right] = \left[A_0, B_0\right] +\boldsymbol{c}\otimes \boldsymbol{h},
	\ene
	where $ \boldsymbol{c} \in \bR^n$. It follows from \eqref{hvector} that $(\overline{A}, \overline{B})\in\Sigma_{\mD(\boldsymbol{v}_1,...,\boldsymbol{v}_p|A_0,B_0)}$.

	For any $i\in \{1,...,\ell\}$, it follows from \eqref{barAB2} that 
	\bee \label{t54}
	\boldsymbol{h}_i'\text{vec}([\overline{A}, \overline{B}]) = \boldsymbol{h}_i'\text{vec}\left(\left[A_0, B_0\right]\right) + \boldsymbol{c}'\cdot\text{vec}^{-1}(\boldsymbol{h_i})\boldsymbol{h}.\ene
	Since $\text{vec}^{-1}(\boldsymbol{h_i})\boldsymbol{h}\neq\boldsymbol{0}, \forall i\in\mathcal{C}_1$ and $\mathcal{S}_i$ is a bounded set, one can choose a sufficiently large vector $\boldsymbol{c}=\boldsymbol{c}_0$ such that $	\boldsymbol{h}_i'\text{vec}([\overline{A}, \overline{B}])\not\in\cS_i$ for all $i\in\cC_1$. 
If $i \in \mathcal{C}_2$, it holds that $\text{vec}^{-1}(\boldsymbol{h_i})\boldsymbol{h}=\boldsymbol{0}$. Then, \eqref{t54} implies that $\boldsymbol{h}_i'\text{vec}([\overline{A}, \overline{B}]) = \boldsymbol{h}_i'\text{vec}\left(\left[A_0, B_0\right]\right)$.  

It only remains to show that under  $\boldsymbol{c}=\boldsymbol{c}_0\in \bR^n$, the system $(\overline{A}, \overline{B})$ from \eqref{barAB2} does not belong to $\Sigma_\mP$, i.e., $\Sigma_{\mD(\boldsymbol{v}_1,...,\boldsymbol{v}_p|A_0,B_0)}\not\subseteq\Sigma_\mP$, which jointly with Lemma \ref{half} completes the proof.  To this purpose, let $\Sigma_i''=\Sigma_i^c, \forall i\in\mathcal{C}_1$ and $\Sigma_j''=\Sigma_j', \forall j\in\mathcal{C}_2$. Then for $\boldsymbol{c}=\boldsymbol{c}_0$ we have 
	\bee \label{constraint_oab}(\overline{A}, \overline{B})\in \cap_{i=1}^\ell \Sigma_i''. \ene
Suppose that Algorithm \ref{algorithm1} terminates in Line 8 when $i=i_0$. Then it is clear that $j\in\mathcal{C}_2$ for any $j>i_0$. Like the case for $(A_0,B_0)$, it holds that $(\overline{A}, \overline{B})\in\Sigma_\mP$ if and only if 
	\bee \label{t56} (\overline{A},\overline{B}) \in \Sigma_1 \odot \Sigma_2\odot \ldots \odot \Sigma_{i_0}.\ene
	Next, if the operator before $\Sigma_{i_0}$ is $\cup$ in \eqref{interunion}, then it follows from the line 5 in Algorithm 1 that $\Sigma_j''=\Sigma_j^c$ for all $j<i_0$. Jointly with that $\Sigma_{i_0}''=\Sigma_{i_0}^c$, we obtain that \eqref{t56} is not satisfied. If the operator before $\Sigma_{i_0}$ is $\cap$ in \eqref{interunion}, then it follows from \eqref{constraint_oab} that $(\overline{A}, \overline{B})\in \Sigma_{i_0}^c$, which implies that \eqref{t56} is not satisfied. Thus, $(\overline{A}, \overline{B})\notin\Sigma_\mP$. 
	
	Suppose that Algorithm \ref{algorithm1} terminates in Line 15. Similarly it follows that $(\overline{A}, \overline{B})\in\Sigma_\mP$ if and only if $(\overline{A}, \overline{B}) \in \Sigma_1$. However, if it terminates in Line 15,  it follows from \eqref{constraint_oab} that $(\overline{A}, \overline{B})\in \Sigma_1^c$, which implies that $(\overline{A}, \overline{B})\notin\Sigma_\mP$.
	
	
	\section{Proof of Remark \ref{brackets}}
	\label{app_brackets}
	
	For the case in Remark \ref{brackets}, we still only need to show the minimality of $\text{im}(M)$. To this end, we follow similar steps with Appendix \ref{sec::proofTheo5}. The key difference is that we decide $\Sigma_i'$ by Algorithm \ref{algorithm2} instead. 
	\begin{algorithm}[t!] 
		\caption{Construction of $\Sigma_i', \forall i\in\{1,\ldots,\ell\}$ for Remark \ref{brackets}}\label{algorithm2}
		\hspace*{0.02in} {\bf Input:}
		$\Sigma_\mP$ in Remark \ref{brackets}.\\
		\hspace*{0.02in} {\bf Output:} 
		$\Sigma_r',r=1,...,\ell.$
		\begin{algorithmic}[1]
			\State Let $L$ denote the expression by which $\Sigma_\mP$ is defined. 
			\Loop
			\State Let $l$ be the number of operators in $L$. 
			\State Sort the operators in $L$ by the order that they are actually executed, and call them in order $\odot_1, \odot_2, \ldots, \odot_{l}$, i.e., $\odot_1$ is first calculated in $L$ and $\odot_{l}$ is the last. 
			\If{one side of $\odot_l$ is $\Sigma_i$ with $i \in \mathcal{C}_1$}
			\State let $\Sigma_i'=\Sigma_i$; 
			\If{$\odot_l$ is $\cup$}
			\State let $\Sigma_j'=\Sigma_j^c$ for all $\Sigma_j$ in the other side; 
			\Else
			\State let $\Sigma_j'=\Sigma_j$ for all $\Sigma_j$ in the other side; 
			\EndIf
			\State \textbf{break}
			\Else
			\State $L$ $\leftarrow$ one side of $L$ that contains a set $\Sigma_i$ with $i \in \mathcal{C}_1$;
			\If{$\odot_l$ is $\cup$}
			\State let $\Sigma_j'=\Sigma_j^c$ for all $\Sigma_j$ in the other side; 
			\Else
			\State let $\Sigma_j'=\Sigma_j$ for all $\Sigma_j$ in the other side.
			\EndIf
			\EndIf
			\EndLoop
		\end{algorithmic}
	\end{algorithm}
	
	\section{Proof of Theorem \ref{info_stru}}\label{append_id}
($\Leftarrow$) 
		For any $(j,i) \in \mathcal{I}_A$, it follows that $i\in \cI_\mP$ and 
		\bee \label{e1} \begin{split}
		a_{ji}^* &= \boldsymbol{e}_{j}'[A^*,B^*]\boldsymbol{e}_{i}
		\\
		&=\left\{[A^*,B^*][X'_-,U'_-]'Q\right\}_{jl}\\
		&=\{X_+Q\}_{jl}\\
		&=0,
		\end{split} \ene
		where $l$ is the column index of $\boldsymbol{e}_i$ in $[\boldsymbol{e}_r, r\in\mathcal{I}_\mP]$. The second equation in \eqref{e1} follows from \eqref{def_Q} and the last equation follows from that $X_+Q\in \cM_{\mP}$ and \eqref{def_Mp}. Similarly, for any $(p,q) \in \mathcal{I}_B$, we have $b_{pq}^*=0$. Jointly with \eqref{stru_sigma_p}, it follows that the system $(A^*,B^*)$ has property $\mP$.
		
		($\Rightarrow$)
		By \eqref{def_Q}, it holds that 
		\bee \begin{split}
			X_+Q &=[A^*,B^*][X'_-,U'_-]'Q \\
			&=[A^*,B^*][\boldsymbol{e}_r, r\in\mathcal{I}_\mP].
		\end{split} \ene
		Since $(A^*,B^*) \in \Sigma_\mP$, it follows directly from \eqref{def_Mp} that $X_+Q\in \cM_{\mP}$.
		
\bibliographystyle{agsm}
\bibliography{mybib}

\begin{wrapfigure}{l}{0.2\linewidth}
	\includegraphics[width=\linewidth]{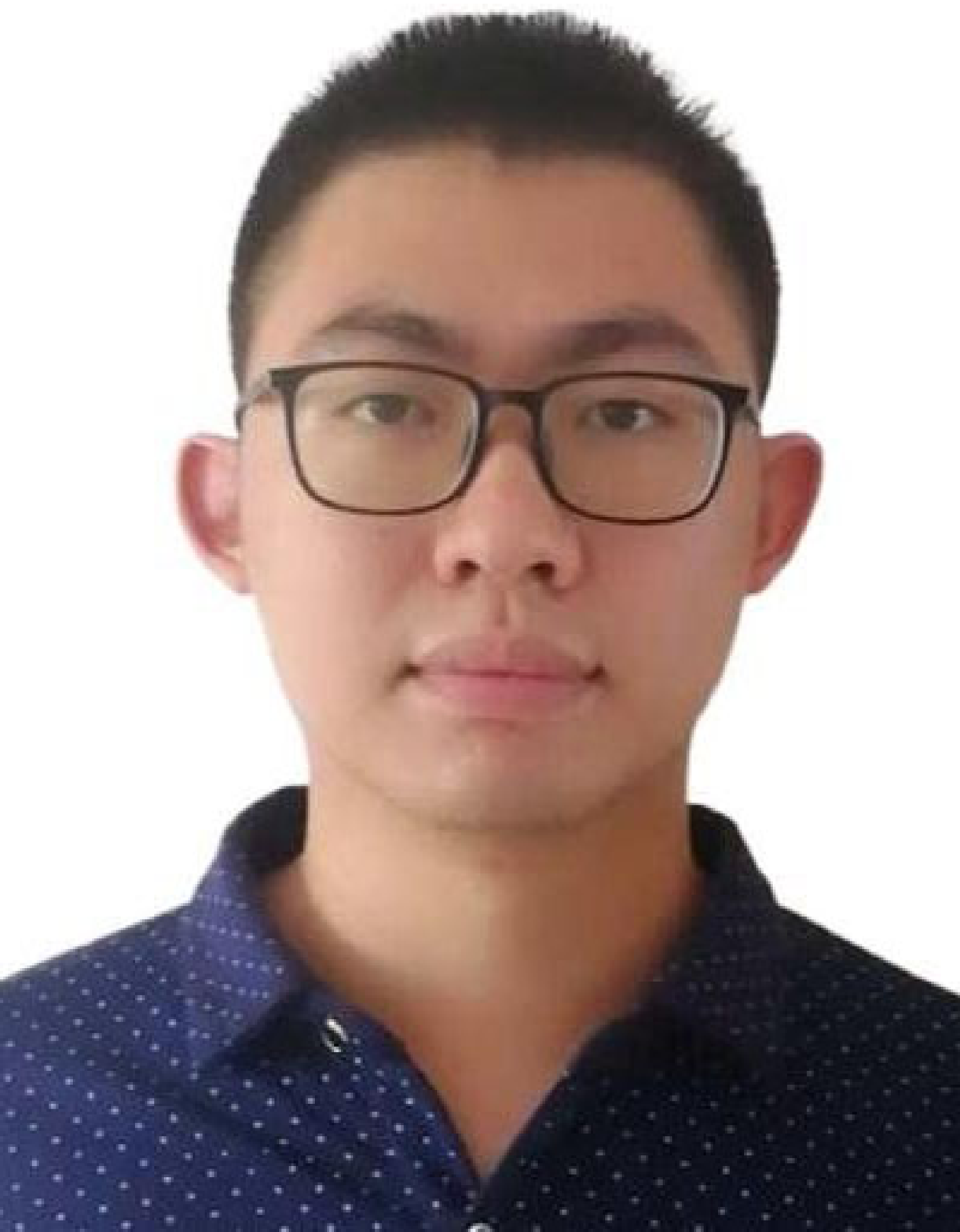}
\end{wrapfigure}
\textbf{Shubo Kang} received the B.S. degree, in 2020, from the Department of Automation, Tsinghua University, Beijing, China, where he is currently working toward the Ph.D. degree. His research interests include data-driven control and reinforcement learning. \\ 

\begin{wrapfigure}{l}{0.2\linewidth}
	\includegraphics[width=\linewidth]{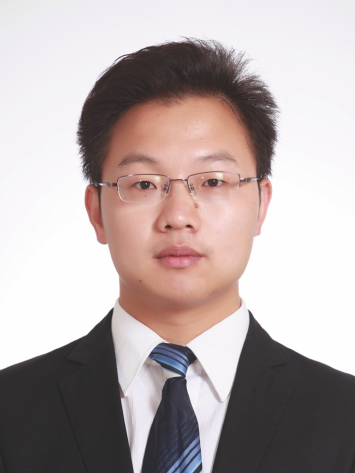}
\end{wrapfigure}
\textbf{Keyou You}(SM'17) received the B.S. degree in Statistical Science from Sun Yat-sen University, Guangzhou, China, in 2007 and the Ph.D. degree in Electrical and Electronic Engineering from Nanyang Technological University (NTU), Singapore, in 2012. After briefly working as a Research Fellow at NTU, he joined Tsinghua University in Beijing, China where he is now a tenured Associate Professor in the Department of Automation. He held visiting positions at Politecnico di Torino,  Hong Kong University of Science and Technology,  University of Melbourne and etc. His current research interests include networked control systems, distributed optimization and learning, and their applications.

Dr. You received the Guan Zhaozhi award at the 29th Chinese Control Conference in 2010 and the ACA (Asian Control Association) Temasek Young Educator Award in 2019. He received the National Science Fund for Excellent Young Scholars in 2017. He is currently an Associate Editor for the IEEE Transactions on Control of Network Systems, IEEE Transactions on Cybernetics,  and Systems \& Control Letters.
	
\end{document}